 \documentclass[final,5p,times,twocolumn]{elsarticle}

\usepackage{float}
\journal{Physics Letters A}

\usepackage{graphicx}
 \usepackage{epsfig}
\begin{document}
\def\ds{\displaystyle}
\begin{frontmatter}

\title{PT Symmetric Floquet Topological Phase in SSH Model}
\author{Z. Turker$^{1}$, S. Tombuloglu$^{2}$, C. Yuce$^{3}$}
\address{ $^{1}$Faculty of Engineering, Near East University, Nicosia, Cyprus\\
 $^{2}$Saglik Hizmetleri MYO, Kirklareli University, Turkey\\
$^{3}$Physics Department, Anadolu University, Eskisehir, Turkey}
\ead{ssemishh@hotmail.com} \fntext[label2]{}
\begin{abstract}
We consider periodically modulated Su-Schrieffer-Heeger (SSH) model with gain and loss. This model, which can be realized with current technology in photonics using waveguides, allows us to study Floquet topological insulating phase. By using Floquet theory, we find the quasi-energy spectrum of this one dimensional PT symmetric topological insulator. We show that stable Floquet topological phase exists in our model provided that oscillation frequency is large and the non-Hermitian degree is below than a critical value. 
\end{abstract}


\end{frontmatter}


\section{Introduction}

The discovery of topological insulators in $2D$ and $3D$ has attracted a great deal of attention in the last decade \cite{hasan}. A topological insulator has gapless robust edge states while its energy spectrum is gapped in the bulk. Robust edge states, which are protected against arbitrary perturbations, appear in time-independent systems since the bulk energy gap closes when the topologically nontrivial system is in contact with a topologically trivial one. We note that topological phase is not restricted to two and three dimensional systems. Of special importance in $1D$ topological insulators is the Su-Schrieffer-Heeger  (SSH) model \cite{andre,AAH,A1}, which is a tight-binding model with alternating hopping amplitudes. On the contrary to $2D$ topological insulators whose edge modes are propagating either chiral or helical modes depending on the topological invariant, edge modes in $1D$ are accumulated at edges and decays rapidly away from edges. \\
Over the past few years, the periodical table of topological insulators was constructed for Hermitian Hamiltonians. Recently, extension of topological phase to non-Hermitian systems has also attracted considerable attention \cite{PTop2,PTop3,ekl56,PTop4,hensch,PTop1}. We note that a non-Hermitian Hamiltonian may admit real spectrum as long as non-Hermitian degree is below than a critical number and the Hamiltonian is $\mathcal{PT}$ symmetric, where $\mathcal{P}$ and $\mathcal{T}$ operators are parity and time reversal operators, respectively. Hu and Hughes \cite{PTop2} and Esaki et. al.  \cite{PTop3} investigated topological phase in some non-Hermitian $\mathcal{PT}$ symmetric system at almost the same time. They concluded that topological phase are not compatible in the $\mathcal{PT}$ symmetric region. In \cite{PTop2}, the authors discussed that non-Hermitian topological phases could only be compatible for systems without Dirac-type Hamiltonians. A non-Hermitian generalizations of the Luttinger Hamiltonian and Kane-Mele model was considered in \cite{PTop3} and it was shown that robust zero energy edge states decay in time because of the imaginary part of energy eigenvalues. Although the initial attempts failed, some other authors continued to find stable topological phase in non-Hermitian systems \cite{ekl56,PTop4,hensch}. But they found unstable topological phases. In 2015, Zhu, Lu and Chen used SSH model with gain and loss \cite{PTop1}. They found that the energy eigenvalues are real valued in topologically trivial region while they become complex valued in topologically nontrivial region. Fortunately, one of us found stable topological phase in a non-Hermitian Aubry-Andre model for the first time in 2015 \cite{cem0001}. A year later, an experiment was realized \cite{sondeney1} and confirmed the theoretical prediction of \cite{cem0001}. In the experiment \cite{sondeney1}, lossy waveguides were used and states that are localized at the interface between two topologically distinct $\mathcal{PT}$-symmetric photonic lattices were observed through fluorescence microscopy. After the first experimental realization, the topic of topologically insulating phase in non-Hermitian systems has attracted great attention. In $1D$, non-Hermitian topological phase was studied by various authors \cite{sbt1,sbt3,sbt6,sbt11,sbt8}. It was shown in \cite{sbt2} that topological insulating phase can also be realized only by gain and loss. Topological phase was also investigated for a generalized non-Hermitian Su-Schrieffer-Heeger model \cite{sbt4}. It is interesting to note that complex Berry phase in non-Hermitian systems was calculated numerically \cite{sbt5}. Chiral topological edge modes in a non-Hermitian variant of the 2D Dirac equation was studied in \cite{sbt100}. Non-Hermitian topological superconductor and Majarona modes have recently been investigated by some authors \cite{sbt12,sbt16,sbt17,sbt7,sbt18}. The topic of non-Hermitian extension of topological phase is still in its infancy and there are some open problems such as bulk-boundary correspondence and topological invariants in non-Hermitian systems.\\
Standard classification of topological insulators and superconductors, which tells us topological invariant for a given Hamiltonian in any dimension by looking at the three discrete symmetries fails if the system is time-dependent. Another kind of topological insulator that appears in time-periodic system is called Floquet topological insulator \cite{flotop1,flotop11,flotop2}. Floquet topological phase has been extensively studied for Hermitian systems. However, there has been few theoretical \cite{sbt10,sbt15} and experimental  \cite{new11,new12,new13} papers in non-Hermitian systems. The theoretical model proposed in \cite{sbt10} includes $z$-dependent gain and loss term in the Hamiltonian. Although Floquet topological phase appears in such a model, it is not experimentally feasible since gain and loss changes periodically with propagation distance. In this Letter, we propose an experimentally feasible theoretical model for the observation of topological Floquet insulator in a non-Hermitian system. Our model can be tested with current technology on photonic lattices.

\section{Model}

The Su-Schrieffer-Heeger (SSH) model is a two band model that describes spinless fermions hopping on a $1D$ tight binding lattice with staggered tunneling amplitudes. This model and Aubry-Andre model \cite{cmyc} are extensively used models in the study of topological phase in one dimension. It is well known that the SSH model has two different topologically distinct dimerized states. Topological zero energy states appears at the interfaces between these two distinct states. In this study, we consider a variant of the SSH model to search for Floquet topological phase in a non-Hermitian system. Let us first discuss the non-Hermitian character of our $1D$ tight binding system. In the experiment \cite{sondeney1}, alternating gain and loss are introduced into the system to make the Hamiltonian non-Hermitian. Here we consider that two non-Hermitian impurities (gain and loss) are arranged at symmetrical sites with respect to the center of the lattice, i.e., particles are injected on the $\ds{j}$-th site and removed from the $\ds{(N-j+1)}$-th site, where $\ds{N}$ is the number of lattice sites. Generally speaking, the main difference between the two cases is that the critical non-Hermitian strength below which the corresponding spectrum is real becomes bigger in our case. To study Floquet topological transition in our non-Hermitian system, we further suppose that the system is modulated periodically. In \cite{flotop1,flotop2}, tunneling amplitude is supposed to change periodically in time to study Floquet topological insulator. Here we use another model. We suppose that potential gradient changes periodically with longitudinal distance. This is a more realistic consideration from the experimental point of view in PT symmetric photonics systems. The Hamiltonian reads
\begin{eqnarray}\label{mcabjs4}
H=-T\sum_{n=1}^{N-1}\left(1+\lambda \cos{(\pi  n+\Phi)}\right)  a^{\dagger}_{n} a_{n+1}+h.c.\nonumber\\
+\sum_{n=1}^{N}f(z)(n-n_0) a^{\dagger}_{n} a_{n}+  i\gamma( a^{\dagger}_j a_j- a^{\dagger}_{N-j+1} a_{N-j+1})
\end{eqnarray}
where $\ds{T}$ is unmodulated tunneling amplitude which is assumed to be real valued, $f(z)$
is the real valued $z$-dependent potential gradient, $\ds{a^{\dagger}_n}$ and $\ds{a_n}$ denote the creation and annihilation operators of particles on site $\ds{n}$, respectively and  the parameter $\gamma$ represents non-Hermitian degree describing the strength of gain/loss material that is assumed to be balanced in the system. The constant $n_0$ determines the zero point of the corresponding term ($n_0=N/2$ for even N and $n_0=(N+1)/2$ for odd N ) and the constant $\lambda$ is the strength of the modulation. The first term in the Hamiltonian corresponds to the usual SSH model and the term with $f(z)$ accounts for the periodical driving. The last term is the non-Hermitian potential due to two non-Hermitian impurities. As it is usual on the topic of topological insulator in $1D$, the modulation phase $\Phi$ is
an additional degree of freedom. We emphasize that an experiment described by the above Hamiltonian can be realized with current technology. Such an experiment, which will be the first one to observe Floquet topological insulator in a non-Hermitian system, can check our findings.\\
Let us first discuss the $\mathcal{PT}$ symmetry of the above Hamiltonian qualitatively. The standard SSH model is Hermitian and $\mathcal{PT}$ symmetric. With the inclusion of non-Hermitian impurities at the symmetrical points, the $\mathcal{PT}$ symmetry is still not lost in the system. This picture generally changes in the presence of the linearly changing modulation. It is easy to see that such a term breaks the translational invariance of the system. Fortunatelly, the $\mathcal{PT}$ symmetry is still restored if $f(z)$ is chosen specifically as follows
\begin{equation}\label{revf241}
f(z)= \kappa~\omega \sin({\omega}z+\phi  )
\end{equation}
where $\omega$ is the modulation
frequency, $\ds{\phi}$ is the initial phase and the constant
$\ds{ \kappa~\omega}$ is the amplitude. As a result, we say that the $\mathcal{P}\mathcal{T}$ symmetry is not broken in our system. So, one expects that the system may have a non-zero threshold of $\gamma$ for spontaneous $\mathcal{P}\mathcal{T}$ symmetry breaking.\\
Although our system is $\mathcal{P}\mathcal{T}$ symmetric, associated topological symmetries are broken in our system. In fact, the term with $f(z)$ breaks the discrete symmetries that the standart SSH model have. As it is well known in the theory of topological insulators, the standart SSH model has time reversal, particle-hole and chiral symmetries. This makes the SSH model BDI class in the standard topological classification. In the SSH model, topological transition occurs at $\ds{\lambda=0}$. More specifically, the SSH model is $Z$ topologically trivial (nontrivial) when $\ds{\lambda>0}$ ($\ds{\lambda<0}$). With the additional term with $f(z)$ into the Hamiltonian, no topological phase appears since the discrete symmetries are broken. In $1D$, a Hamiltonian without these three discrete symmetries is topologically nontrivial according to the standard classification of topological insulator. However, we can instead study Floquet topological phase. It is well known that there are some time periodical systems where the system has no topological phase but Floquet topological phase. More specifically, Floquet topological phase appears in a system if not the original system but the effective system has desired discrete symmetries. \\
To look for Floquet topological phase in our non-Hermitian system, we should find the corresponding effective Hamiltonian in high frequency regime. Below we first perform analytical calculation and find the corresponding effective Hamiltonian. After getting some analytical notion in the high frequency limit, we will then perform numerical computation. We will numerically show that topological phase is not possible in the low frequency limit. \\
Consider the Hamiltonian (\ref{mcabjs4}). A common way in the studies of high frequency Floquet systems is to find a time-independent effective Hamiltonian. As it is discussed in \cite{effective01}, the tunneling parameter is replaced by an effective tunneling
amplitude, $\ds{{T_{eff.}} }$ in the high-frequency domain. With
application of the high-frequency Floquet approach, the
Hamiltonian (\ref{mcabjs4}) can then effectively be described as
\begin{eqnarray}\label{ameffh}
H_{eff.}= -T_{eff.}\sum_{n=1}^N\left(1+\lambda \cos{(\pi  n+\Phi)}\right) 
a^{\dagger}_{n} a_{n+1}+h.c.\nonumber\\
+\sum_{n=1}^{N}  i\gamma( a^{\dagger}_j a_j- a^{\dagger}_{N-j+1} a_{N-j+1})
\end{eqnarray}
where the effective tunneling is given by
\begin{eqnarray}
T_{eff.}=T\overline{ \int_0^z e^{i\eta}dz^{\prime}}=T \mathcal{J}_{0} (\kappa)
\end{eqnarray}
where overline denotes the average over $z$ and $\eta$ is given by
$\ds{\eta(z)=\int_0^{z}{f(z^{\prime})~dz^{\prime}}}$. Here, we have used a useful expansion of the
oscillatory term $\ds{e^{i\eta}}$ in terms of Bessel functions by
using the Jacobi-Anger expansion; $\ds{
e^{i\kappa\sin(x)}=\sum_{m} \mathcal{J}_m ( \kappa )e^{imx} }$,
where $\mathcal{J}_m$ is the
$m$-th order Bessel function of first kind. Note that the periodical $z$-dependent term is absent in the effective Hamiltonian. Since the effective Hamiltonian is $z$-independent, one would use standart classification of topological insulators to study the existence of topological phase in the effective system if the Hamiltonian is Hermitian.\\
If we compare the original and effective Hamiltonians, we say that the presence of the monochromatic modulation
corresponds to a modification of the tunneling amplitude at the expense of absence of the $z$-dependent modulating term in the original Hamiltonian. The scaling factor is given by Bessel function. In the absence of z-dependent modulation, $\kappa=0$, the
Bessel function takes its maximum value, $\ds{ \mathcal{J}_{0}(0)=1}$. We note that the Bessel function $\ds{ \mathcal{J}_{0} (\kappa)}$ is roughly like an oscillating sine function that decays proportionally as $\kappa$
increases.  At some certain values of $\ds{\kappa}$, the Bessel function becomes equal to zero. In other words, no particle can tunnel to the neighboring sites. This effect is known as the dynamical localization effect since tunneling is suppressed dynamically. The first zeros of the Bessel function happens at $\ds{\kappa=2.405}$. At this value, our system admits complex energy spectrum. If $\ds{\kappa}$ is increased above from this value, then tunneling is partially restored and the spectrum becomes real as long as the non-Hermitian degree is below than a critical number, $\ds{\gamma< \gamma_{PT}}$.\\
The above approach, which is valid only in high frequency regime is the basis of Floquet topological phase. The original Hamiltonian (\ref{mcabjs4}) does not have topological phase since the discrete symmetries are broken. However, the effective Hamiltonian (\ref{ameffh}) in the high frequency regime is just the SSH model with gain and loss. As it was discussed in our previous paper \cite{cem0001}, there exists topological zero energy modes for this effective Hamiltonian. An experiment on the SSH model with gain and loss was also realized and topological modes were observed in \cite{sondeney1}. The region of nontrivial topological phase depends on whether the number of lattice sites is odd or even. More specifically, zero energy edge states appear for all $\Phi$ (for all $\Phi$ except $ 
\pi/2 < \Phi < 3\pi/2$) if $N$ is odd (even). We note that these  zero states are localized around the two edges of the system. It was shown that this is true even when two non-Hermitian impurities are introduced into the system \cite{cem0001}. The reality of the energy spectrum depends dramatically on the position of non-Hermitian impurities. If they are placed exactly at the edges, then the critical non-hermitian degree, $\ds{\gamma_{PT}}$ is exactly equal to zero. If they are placed away from the edges, then $\ds{\gamma_{PT}}$ takes finite values. We note that $\ds{\gamma_{PT}}$ becomes maximum when the  non-Hermitian impurities are placed at the neighboring sites of edges, $\ds{j=2}$. As a result, we say that not the original Hamiltonian but the effective Floquet Hamiltonian has topological phase, which appears in the high frequency limit. The main finding of this paper is that our system has nontrivial Floquet topological edge states in the un-broken $\mathcal{P}\mathcal{T}$ symmetric region. So far, we have analyzed our system analytically in the high frequency regime. Below, we will explore the energy spectrum numerically in the full regime and discuss the validity of our analytical approach. We will also discuss topological transition in our system.\\
\begin{figure}[t]\label{figcem}
{\includegraphics[width=4.5cm]{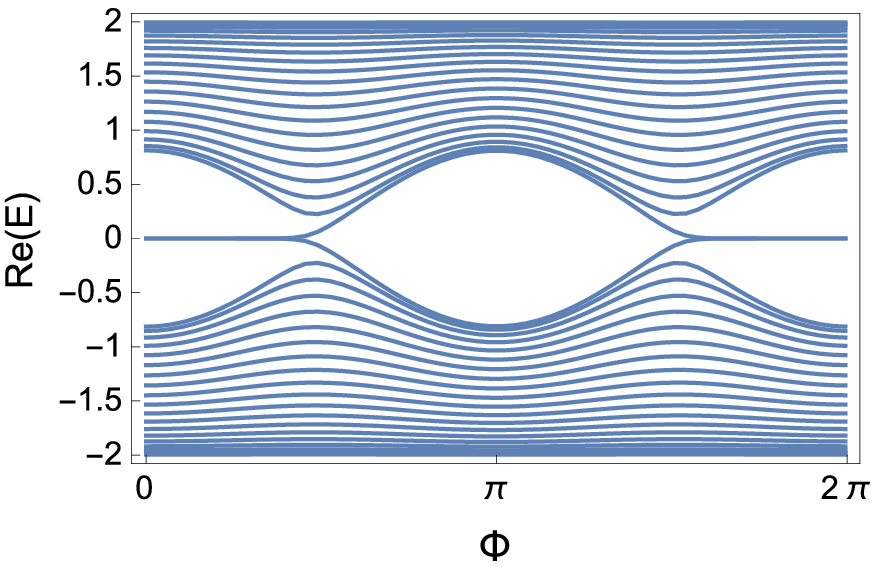}\includegraphics[width=4.5cm]{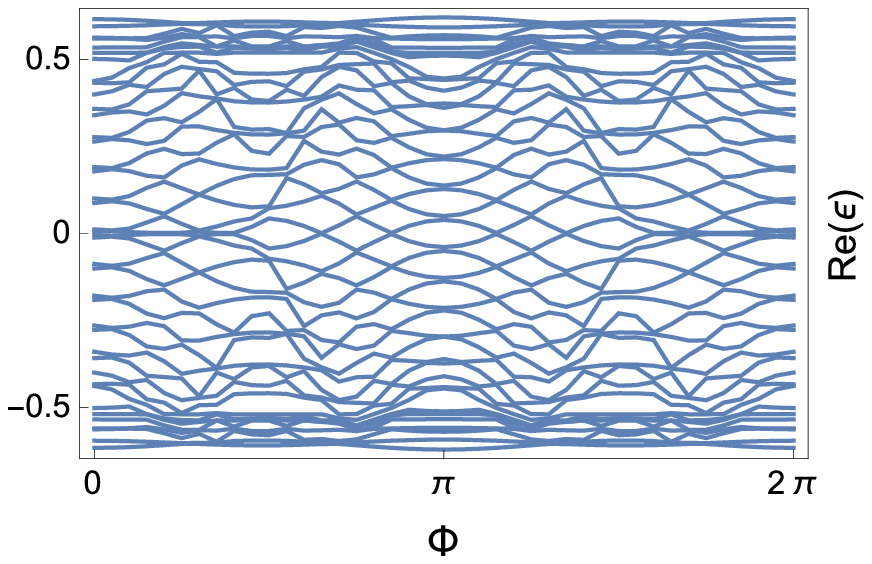}}
{\includegraphics[width=4.5cm]{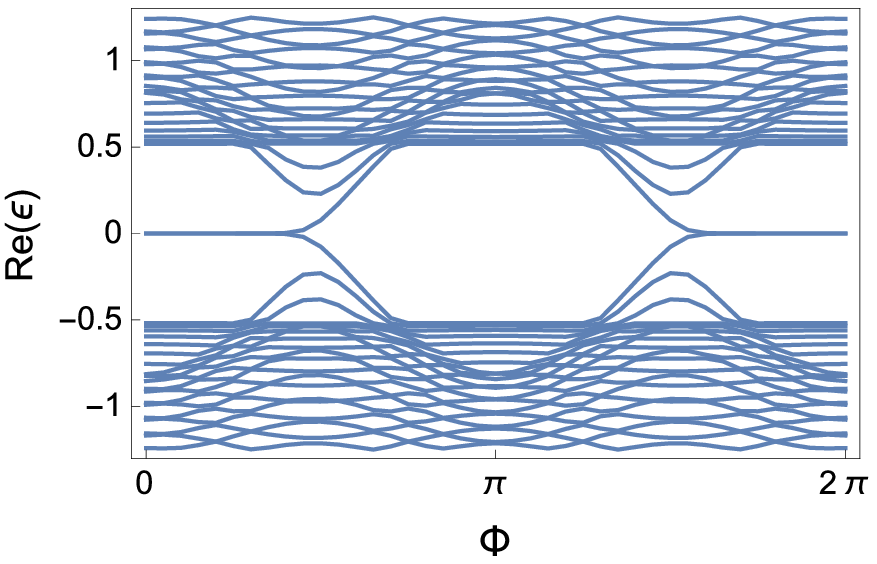}\includegraphics[width=4.5cm]{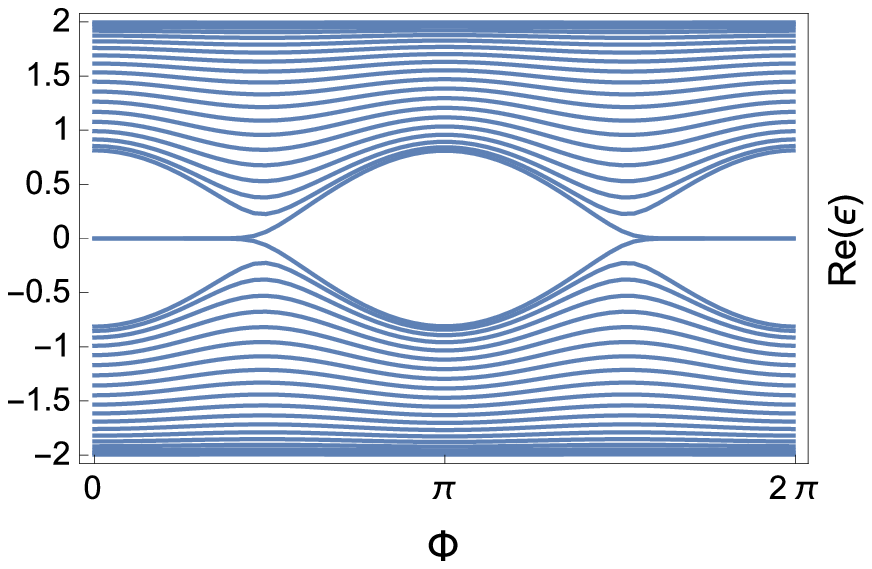}}
\caption{Numerical calculations of the real part of the energy and quasi-energy spectra for the parameters $T=1$, $\ds{\lambda=0.4}$, $\ds{\kappa~\omega=0.05}$, $\gamma=0.2$, $j=2$ and $N=40$ sites as a function of $\Phi$. The top figure at the left is for z-independent case with $\kappa=0$, while the one at the right is for a small modulation frequency $\ds{\omega=0.2\pi}$. The lower two figures at left and right are for an intermediate and a high value of the modulation frequencies, respectively: $\ds{\omega=0.8\pi}$ (left) and $\ds{\omega=45\pi}$ (right). As can be seen, topological zero energy modes exist for $\kappa=0$. The $\mathcal{P}\mathcal{T}$ symmetry is spontaneously broken for small and intermediate values of the modulation frequency. Therefore the spectrum becomes complex valued. The $\mathcal{P}\mathcal{T}$  symmetry is restored and stable Floquet topological edge states appear for large values of $\omega$. }
\end{figure}
It is instructive to briefly introduce the basic ideas of the Floquet formalism used to deal with time dependent ($z$-dependent in our case) periodic Hamiltonians. Consider a non-Hermitian periodical Hamiltonian in the propagation direction with the period $T$, $\ds{H(z)=H(z+T)}$. The energy spectrum of a periodical Hamiltonian can be found using the Floquet theory. According to the Floquet theorem, there exists solutions of the form $\ds{\psi_{\alpha}=e^{-i\epsilon_{\alpha}{z}}~\Phi_{\alpha}}$, where $\Phi_{\alpha}(z)$ is the periodical Floquet eigenstate and the $z$-independent eigenvalues $\ds{\epsilon_{\alpha}}$ are the Floquet quasi energies. Note that the quasi energies are constants. The Floquet spectrum is  $2\pi/T$ periodic in quasi energy just as a repeated zone scheme in quasi momentum for conventional band structure. The dynamics of the periodically driven system is obtained by solving the Floquet-Schrodinger equation $\ds{H_F\Phi_{\alpha}=\epsilon\Phi_{\alpha}}$, where we define the Floquet Hamiltonian as $\ds{H_F=H-i \partial/\partial_z}$. The Floquet states are periodical so they satisfy $\Phi_{\alpha n}=\Phi_{\alpha}e^{in\omega z}$, where orthonormality condition reads $<<\Phi_{\alpha n}|\Phi_{\beta m}>>=1/T\int_0^T<\Phi_{\alpha n}|\Phi_{\beta m}>dz=\delta_{\alpha\beta}\delta_{mn}$ and the second bra-ket notation here is used to denote the integration over the propagation direction $z$. Of particular interest for a non-Hermitian periodical Hamiltonian is the situation when all quasi energies are real. We will numerically find quasi energy spectrum for our non-Hermitian Hamiltonian. In practice, a truncated Floquet Hamiltonian is used in our numerical computation. The corresponding matrix is a $2(2N_F+1)\times2(2N_F+1)$ matrix, where $N_F$ is large enough that the result doesn't depend on $N_F$. Below we present our result for various values of the frequency $\omega$.\\
Let us first review the $z$-independent case with $\kappa=0$. In this case, the system has two bands and zero energy states appear. Consider first the Hermitian limit, $\gamma=0$. In this case, the system has maximum band gap at $\Phi=0$. The band gap closes and reopens as $\Phi$ is varied. Remarkably, zero energy states appear in the spectrum if either $\ds{0<\Phi<\pi/2}$ or $\ds{3\pi/2<\Phi<2\pi}$ when the total site number is even as can be seen from the figure. Note that they would appear in the whole region of $\Phi$ if $N$ is odd. It is well known that these topological zero-energy states are localized around the two edges of the system. Consider now the presence of non-Hermitian term in the Hamiltonian, $\ds{\gamma\neq0}$. Topological zero energy states exist even in the presence of the gain and loss. The corresponding energy eigenvalues are real for $z$-independent case $\kappa=0$ as long as $\ds{\gamma}$ is smaller than a critical value $\ds{\gamma_{PT}}$, which depends on the parameters in the Hamiltonian and decreases with the site number $N$. The $z$-dependent term with $\ds{\kappa\neq0}$ breaks the $\mathcal{PT}$ symmetry spontaneously. The corresponding quasi-energy eigenvalues become completely complex valued if the modulation frequency is small. At intermediate values of the modulation frequency, they are partially complex valued. For large values of $\ds{\omega}$, the $\mathcal{PT}$ symmetry is restored and the spectrum becomes real valued again. As can be seen from the top-right plot in the Fig.1, the energy band changes dramatically and stable topological zero energy modes are lost for small values of $\omega$. This is because of the fact that discrete symmetries are broken with the $z$-dependent term in the Hamiltonian and chiral topological phase no longer exists in the system. As can also be seen from the figure, the spectrum is highly complicated and standard band gap structure is lost. The lower-left figure plots the spectrum for an intermediate value $\omega=0.8 \pi$. To observe Floquet topological edge states, which are immune to disorder, we perform one more numerical computation at $\omega=4 \pi$. We are now in high frequency regime and can confirm our theoretical prediction. Since we choose a small value of $\kappa$, we get $\ds{ \mathcal{J}_0 ( \kappa=0.1/4\pi )=0.999}$. This means that the effective tunneling amplitude is almost equal to original tunneling amplitude. In other words, the spectrum of unmodulated case and the quasi-energy spectrum of modulated case almost coincide. This can be seen from the top-left ($\kappa=0$) the lower-right ($\omega=4\pi$) plots in the figure-1. As can be seen, they are almost identical and both admit zero energy states. This shows that the numerical result is in agreement with our analytical prediction. The Floquet topological zero energy states are stable as long as $\gamma$ is below than a critical number and non-Hermitian impurities are not located exactly at edges in a lattice with even number of total lattice sites. We emphasize that $\mathcal{PT}$ symmetry is automatically broken if $N$ is odd and hence the spectrum is always complex valued for $\gamma\neq0$. The broken $\mathcal{PT}$ symmetry can easily be seen if one considers that the lattice sites at the left and right edges couple to their neighboring sites with different tunneling amplitudes if $N$ is odd.\\
To conclude, we have studied $1D$ tight binding lattice with gain and loss. Our model is a variant of SSH model and can be realized with current technology in photonics using waveguides. We have shown that stable Floquet topological phase exists in our model provided oscillation frequency $\omega$ is large and the non-Hermitian degree $\gamma$ is below than a critical value. Our system is a candidate for the experimental realization of Floquet topological phase in non-Hermitian photonics system. \\
This study is supported by Anadolu University Scientific Research Projects Commission under the grant no: 11705F208

\end{document}